\documentstyle[12pt,aaspp4]{article}
\input{epsf}
\begin{document}
\def\msun{\hbox{${\cal{M}}_{\odot}$}}
\def\massA{\hbox{${\cal{M}}_A$}}
\def\massB{\hbox{${\cal{M}}_B$}}
\def\massAB{\hbox{${\cal{M}}_{A+B}$}}

\title{Speckle Interferometry at the U.S. Naval Observatory. XXIII.}

\author{Brian D.\ Mason, William I.\ Hartkopf\altaffilmark{1}, Sean E.\ Urban\altaffilmark{1}, Jordan D.\ Josties\altaffilmark{2}}
\affil{U.S. Naval Observatory \\
3450 Massachusetts Avenue, NW, Washington, DC, 20392-5420 \\
Electronic mail: brian.d.mason@navy.mil}

\altaffiltext{1}{Retired.}
\altaffiltext{2}{SEAP Intern.}

\begin{abstract}

The results of 3,989 intensified CCD observations of double stars, 
made with the 26-inch refractor of the U.S. Naval Observatory, are 
presented. Each observation of a system represents a combination of 
over two thousand short-exposure images. These observations are 
averaged into 1,911 mean relative positions and range in separation 
from 0\farcs289 to 128\farcs638, with a median separation of 
8\farcs669. Four orbits are improved. This is the 23$^{rd}$ in this
series of papers and covers the period 4 January 2017 through 13 
September 2017.
\end{abstract}


\section{Introduction}

This is the 23$^{rd}$ in a series of papers from the U.S. Naval 
Observatory's speckle interferometry program, presenting results of 
observations obtained at the USNO 26-inch telescope in Washington, 
DC (see, most recently, Mason \& Hartkopf 2017a). 

From 4 January through 13 September 2017, the 26-inch telescope 
was used on 60 of 184 (33\%) scheduled nights. While most nights 
were lost due to weather conditions, time was also lost due to 
testing and upgrades of instrumentation and software, other 
mechanical or software issues, and to a lack of observing personnel.
Instrumentation and the observing technique were as described in 
Mason \& Hartkopf (2017a). Observing was suspended in mid-September
when upgrades to the motors and encoders began. After initial success
in automation seen in this and recent previous entries of this 
series, a more ambitious automation project was initiated in 
September. This will be described in greater detail in the next 
entry in this series.

Individual nightly totals varied substantially, from 7 to 146 
observations per night (mean 66.5). The results yielded 3989
observations (pointings of the telescope) and 3862 resolutions. 
After removing marginal observations, calibration data, tests, and 
$``$questionable measures" a total of 3333 measurements remained. 
These $``$questionable measures" are not all of inferior quality but
may represent significant differences from the last measure, often 
made many decades ago. Before these measures are published they will
need to be confirmed in a new observing season to account for any 
possible pointing or other identification problems. The tabulated 
list of these is retained internally and forms a $``$high priority 
observing list" for subsequent observing seasons. These 3333 
measures were grouped into 1911 mean relative positions. 


\section{Results}

Our 2017 observing list remained the same as the previous, discussed
in Mason \& Hartkopf (2017a). On a given night a pair may be 
observed multiple times in different data collection modes and with 
different magnification as it is not always obvious which will 
produce the best result. Further, as object acquisition is the 
most time-consuming portion of the duty cycle, adding additional 
observations is less consequential. For those intranightly 
observations ($n~=~832$) the rms values are quite low: 
$d\theta~=~0\fdg10$ and $\frac{d\rho}{\rho}~=~0.0020$. A smaller 
number ($n~=~262$) comprise those objects which appear to be slow 
moving\footnote{We assume $\Delta\theta~=~\Delta\rho~=~0$ for 
these.} and were observed on multiple nights. For those internightly
observations the rms values are twice the intranightly values: 
$d\theta~=~0\fdg14$ and $\frac{d\rho}{\rho}~=~0.0044$. We take these
values as representative of the true error.

\subsection{New Pair}

Table 1 presents coordinates and magnitude information from 
CDS\footnote{magnitude information is from one of the catalogs 
queried in the {\it Aladin Sky Atlas}, operated at CDS, Strasbourg, 
France. See {\tt http://aladin.u-strasbg.fr/aladin.gml}.} for a 
pair which is presented here for the first time. It is a closer 
component to a known system. Column one gives the coordinates of 
the primary of the pair. Column two is the WDS identifier while 
Column three is the discoverer designation associated with the known
pair which is used here for the new component as well. Columns four
and five give the visual magnitudes of the primary and secondary, 
and Column six notes the circumstance of the discovery. The mean 
double star positions of our 26$''$ measures (T, $\theta$, and 
$\rho$) of this system is given in Table 3. 

As this pair is quite wide we are able to provide two additional 
measures of relative astrometry from other catalogs using the same 
methodology as described in Wycoff et al.\ (2006) and Hartkopf et 
al.\ (2013). In Table 2 the first two columns identify the system 
by providing its epoch-2000 coordinates and discovery designation 
(as given in Table 1). Columns three through five give the epoch 
of observation (expressed as a fractional Julian year), the 
position angle (in degrees), and the separation (in seconds of 
arc). Note that in all tables the position angle, measured from 
North through East, has not been corrected for precession, and is 
thus based on the equinox for the epoch of observation\footnote{This
has been the standard for double star relative astrometry for several
hundred years and is in accordance with IAU Resolutions from the 
Commissions governing double stars, most recently, Mason et al.\ 
(2016). See \S 4.1.1.}. Columns six and seven provide the source of 
the measure and either a reference or note to the source.

\subsection{Measures of Known Pairs}

Tables 3 and 4 present the relative measurements of double stars 
made with the 26$''$ telescope. Table 3 presents those with no 
calculation for motion, either orbital or linear. As in Table 1, the
first two columns identify the system by providing its epoch-2000 
coordinates and discovery designation. Columns three and four give 
the epoch of observation (expressed as a fractional Julian year) and
the position angle (in degrees). Column five gives the position 
angle error. This is the internightly rms value if one is available 
or the mean value of 0\fdg1 if it is not. Columns six and seven 
provide the separation (in seconds of arc) and its error. As above, 
the error is its internightly value or the mean error 
($d\rho~=~0.0044\rho$). Column eight is the number of nights in the mean 
position. When this is $``$1" the errors in Columns five and seven 
are the mean results as described above. Finally, Column nine is 
reserved for notes. One of the pairs listed in Table 3 is the pair 
listed in Tables 1 and 2 which has not been measured before. Five
pairs, designated with a $``$C" code, are confirmed here for the
first time. Eight very wide pairings, designated with a $``$V" code,
in mulitple systems have their positions determined from vector 
addition. Two pairs have not been measured in over fifty years.
Those are WDS10536$-$0742 = J\phm{88888}90BC, last measured in 1954
(Harshaw 2013) and WDS17128$+$2433 = POU3264, last measured in 1892
(Pourteau 1933).

The 1726 measures presented in Table 3 have a mean separation of 
13\farcs874 and a median value of 9\farcs156. The mean number of 
years since the pair was last observed is 7.05. 

Table 4 presents measurements of doubles where some prediction of
position (orbital or linear) is available. The first eight columns 
are the same as Table 3 above. Columns nine and ten provide the 
O$-$C residual to the determination referenced in Column eleven. 
The final column, like that of Table 3, provides notes. In some 
cases a measure has residuals to more than one calculation. In some 
of those cases the second calculation refers to a new orbit (Table 
5) or linear solution (Table 6) which is described below. 

Not surprisingly, the objects in Table 4 are both closer and more
frequently observed than those of Table 3. The 185 measures 
presented in Table 4 have a mean separation of 11\farcs507 and a 
median value of 4\farcs158. The mean number of years since the pair 
was last observed is 2.88. 

\subsection{Improved Orbits}

Four systems with sufficient data to improve their orbits are 
presented in Table 5 and Figure 1. All of the individual measures 
were weighted by the procedures of Hartkopf et al.\ (2001) and 
calculated with the venerable $``$grid-search" method of Hartkopf et
al.\ (1989).

Table 5 is broken into two groups. The first orbit we characterize 
as $``$improved but still provisional" and is given without errors. 
They fit the data better than the earlier orbit and should give 
reasonable ephemerides over the next several decades, but the 
elements will all require correction over the course of a complete 
orbit before they can be considered even approximately correct. As 
in earlier tables, the first two columns identify the system by 
providing its epoch-2000 coordinates and discovery designation. 
Columns three through nine provides the seven Campbell elements: 
the period (P in years), the semimajor axis (a$''$ in arcseconds), 
the inclination (i) and longitude of the node ($\Omega$), both in 
degrees, the epoch of the most recent periastron passage (T$_o$ in 
years), the eccentricity (e) and the longitude of periastron 
($\omega$ in degrees). Column ten gives the reference to the 
previous $``$best" orbit and Column eleven the orbital $``$grade" 
following the procedures of Hartkopf et al.\ (2001). 

In the second part of Table 5 are the two orbits we characterize 
as $``$reliable", all with shorter periods than those in the
first group. All eleven columns are the same as the first part of 
the table, however, here under each element is its formal error. The
precision of the element is defined by the precision of its error. 
Relative visual orbits of all four systems are plotted in Figure 1,
with the x and y axes indicating the scale in arcseconds. Each solid
curve represents the newly determined orbital elements presented in 
Table 5 and the dashed curve is the orbit of the earlier orbit 
referenced in Column ten.

\begin{figure}[!ht]
\begin{center}
{\epsfxsize 3.2in \epsffile{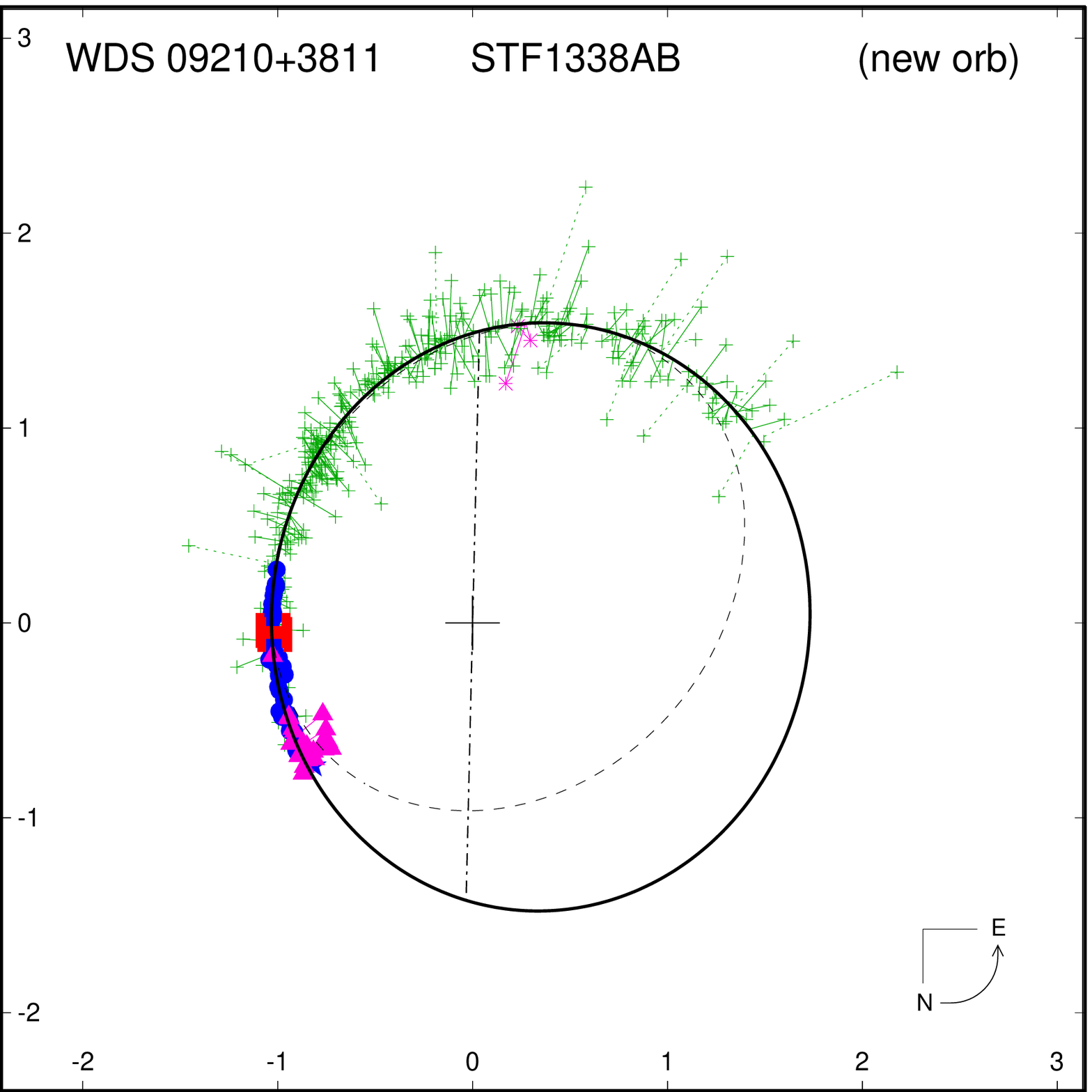} \epsfxsize 3.2in \epsffile{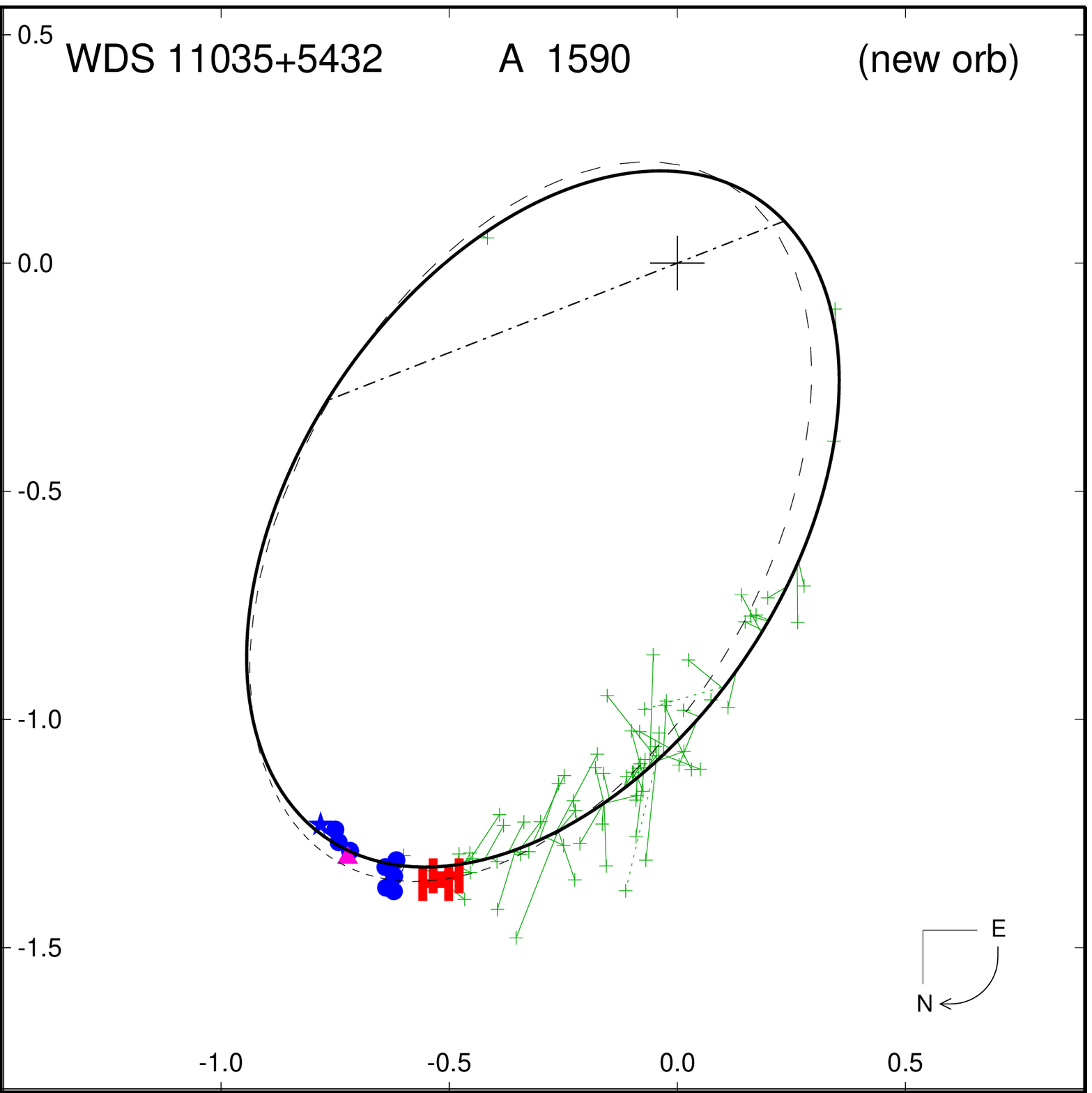}}
{\epsfxsize 3.2in \epsffile{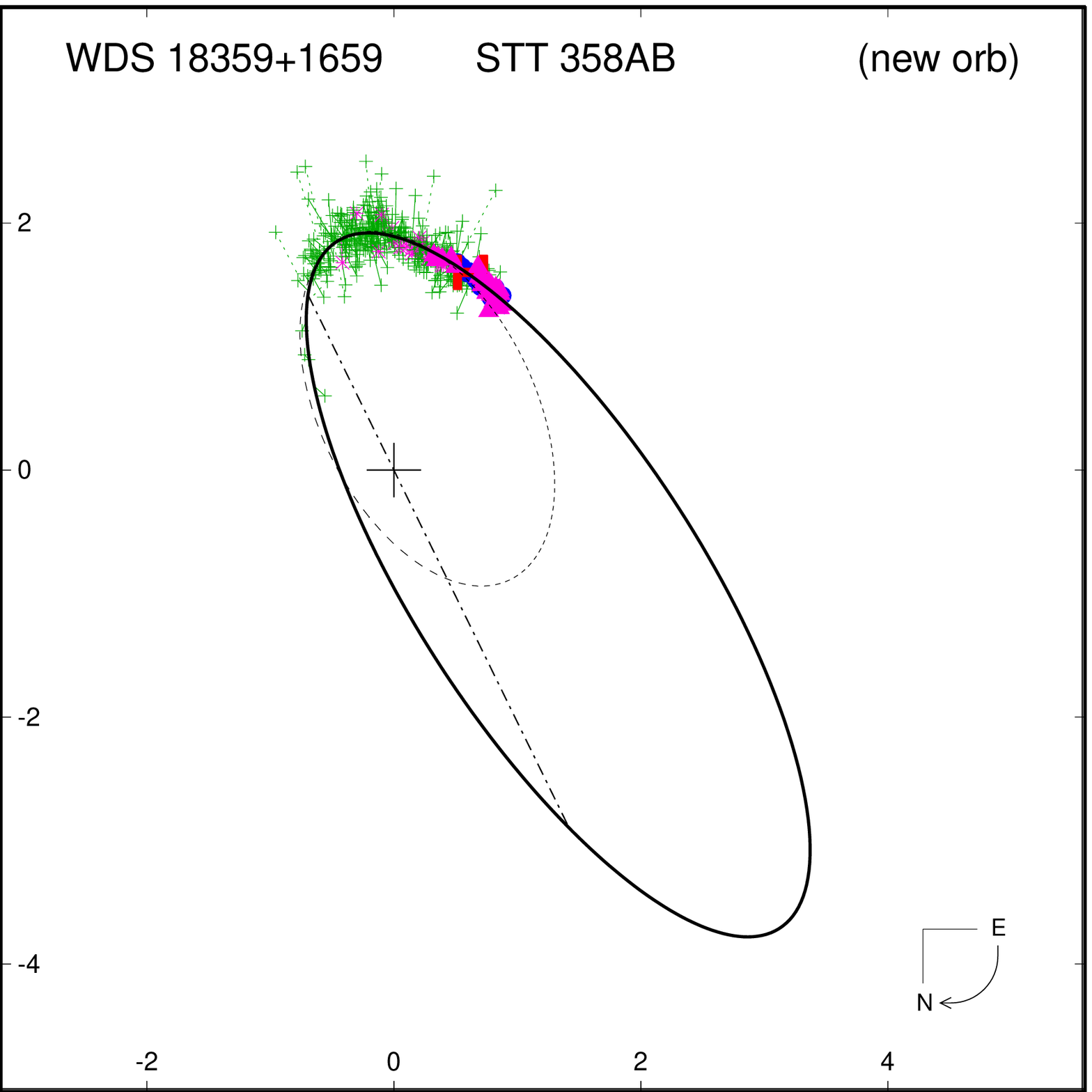} \epsfxsize 3.2in \epsffile{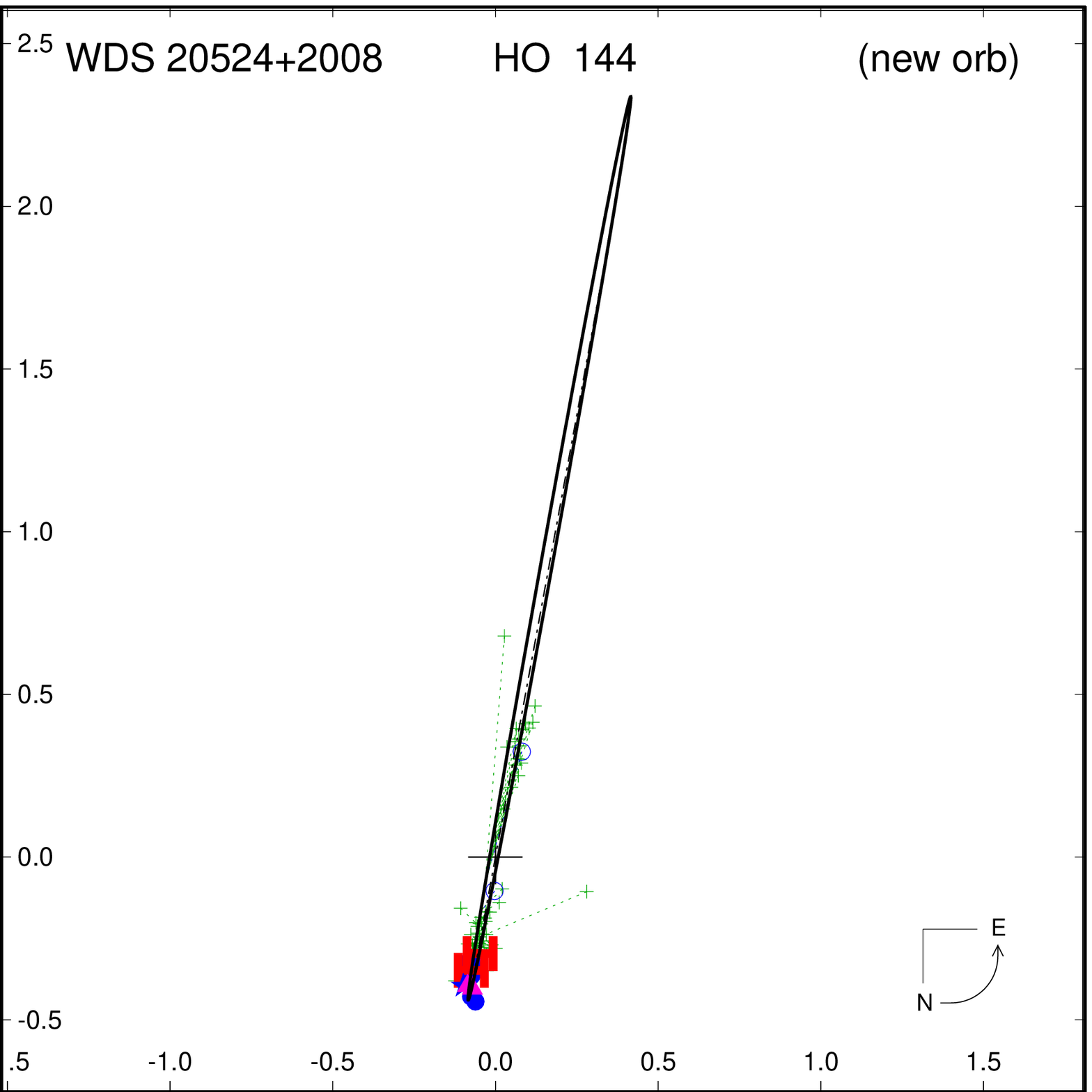}}
\end{center}
\caption{\small Figure 1 illustrates the new orbital solutions, 
plotted together with all published data in the WDS database as well
as the new data in Table 4. In each of these figures, micrometric 
observations are indicated by plus signs, interferometric measures 
by filled circles, conventional CCD by pink triangles, space-based 
measures are indicated by the letter `H', new measure from Table 4 
are plotted as a filled star. ``$O-C$" lines connect each measure to 
its predicted position along the new orbit (shown as a thick solid 
line). Dashed ``$O-C$" lines indicate measures given zero weight in 
the final solution. A dot-dash line indicates the line of nodes, and 
a curved arrow in the lower right corner of each figure indicates the 
direction of orbital motion. The earlier orbit referenced in Table 5 
is shown as a dashed ellipse.}
\end{figure}

\subsection{New Linear Solutions}

Inspection of all observed pairs with either a 30$^{\circ}$ change 
in their relative position angles or a 30\% change in separations 
since the first observation cataloged in the WDS revealed six pairs
whose motion seemed linear. These apparent linear relative motions
suggest that these pairs are either composed of physically 
unrelated stars or have very long orbital periods. Linear elements 
to these doubles are given in Table 6, where Columns one and two 
give the WDS and discoverer designations and Columns three to nine 
list the seven linear elements: x$_{0}$ (zero point in x, in 
arcseconds), a$_{x}$ (slope in x, in $''$/yr), y$_{0}$ (zero point 
in y, in arcseconds), a$_{y}$ (slope in y, in $''$/yr), T$_{0}$ 
(time of closest apparent separation, in years), $\rho_{0}$ (closest
apparent separation, in arcseconds), and $\theta_{0}$ (position 
angle at T$_{0}$, in degrees). See Hartkopf \& Mason (2017) for a 
description of all terms. 

Figure 2 illustrates these new linear solutions, plotted together 
with all published data in the WDS database, as well as the 
previously unpublished data from Table 4. Symbols are the same as in 
Figure 1. In the case of linear plots, the dashed line indicates the 
time of closest apparent separation. As in Figure 1, the direction 
of motion is indicated at lower right of each figure. As the plots 
and solutions are all relative, the proper motion ($\mu$) difference
is assumed to be zero. In some cases, cataloged proper motion 
differences between the components is plotted as a red line.

\begin{figure}[p]
~\vskip -1.8in
\begin{center}
{\epsfxsize 2.8in \epsffile{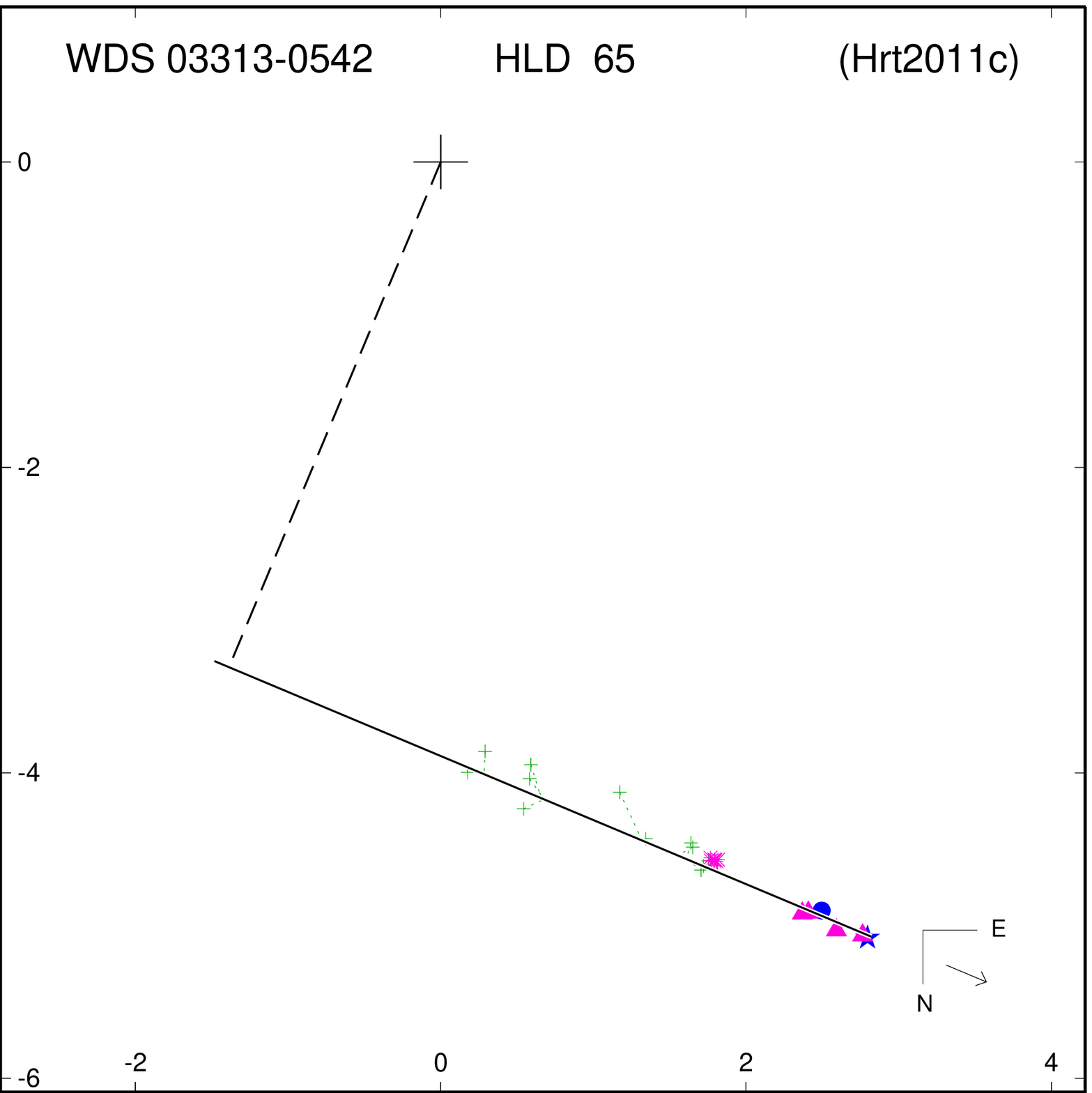} \epsfxsize 2.8in \epsffile{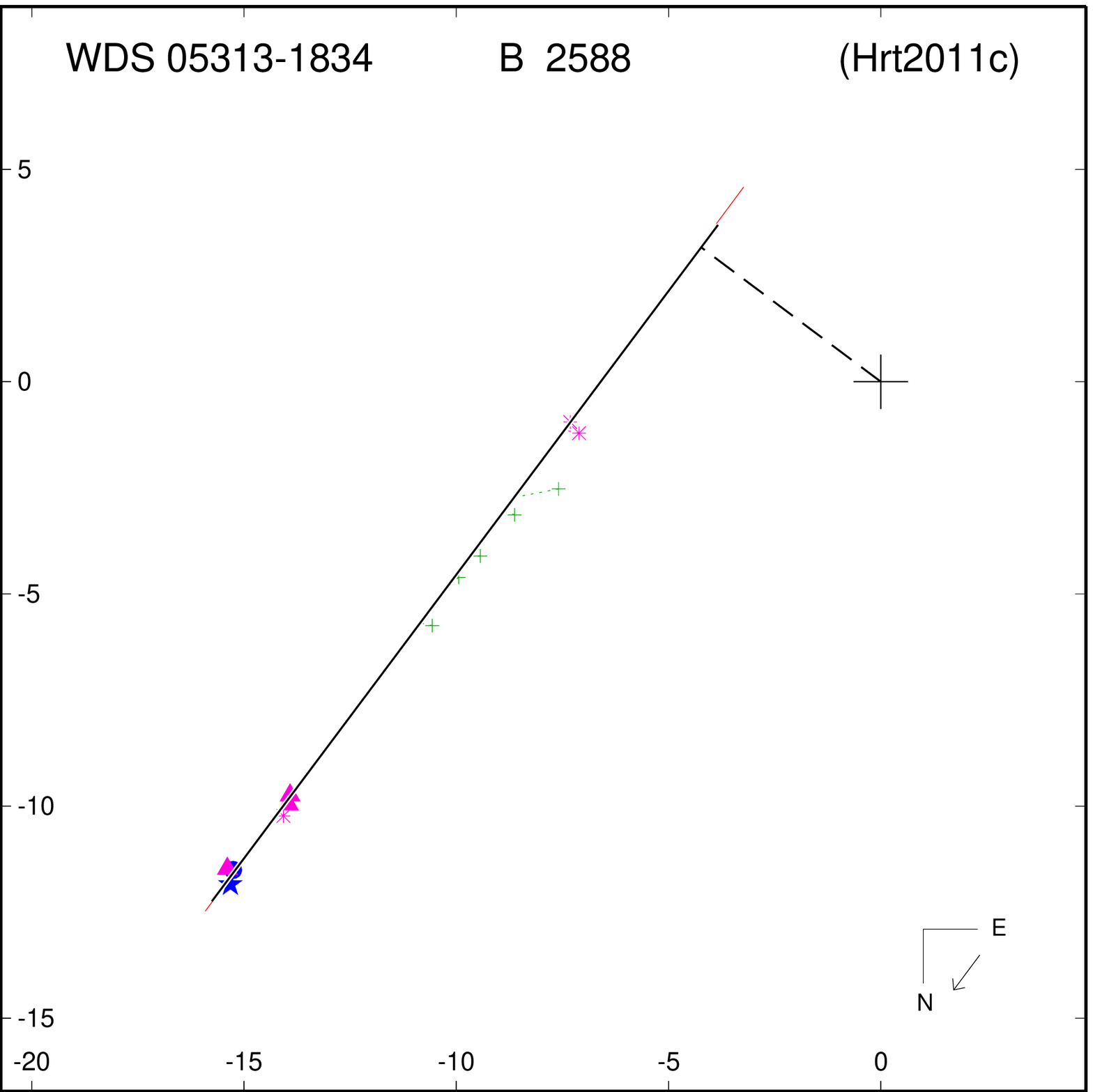}}
\vskip 0.05in
{\epsfxsize 2.8in \epsffile{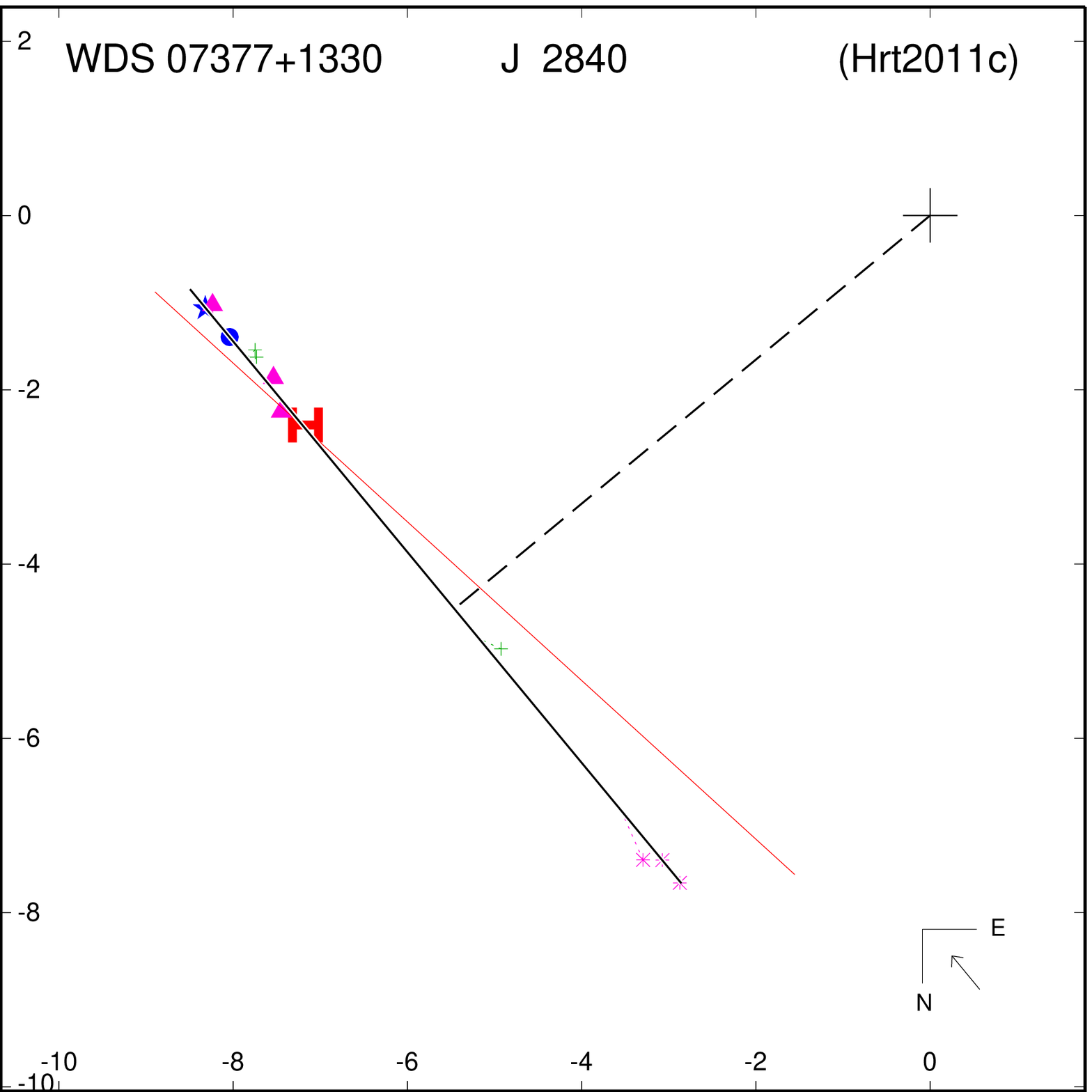} \epsfxsize 2.8in \epsffile{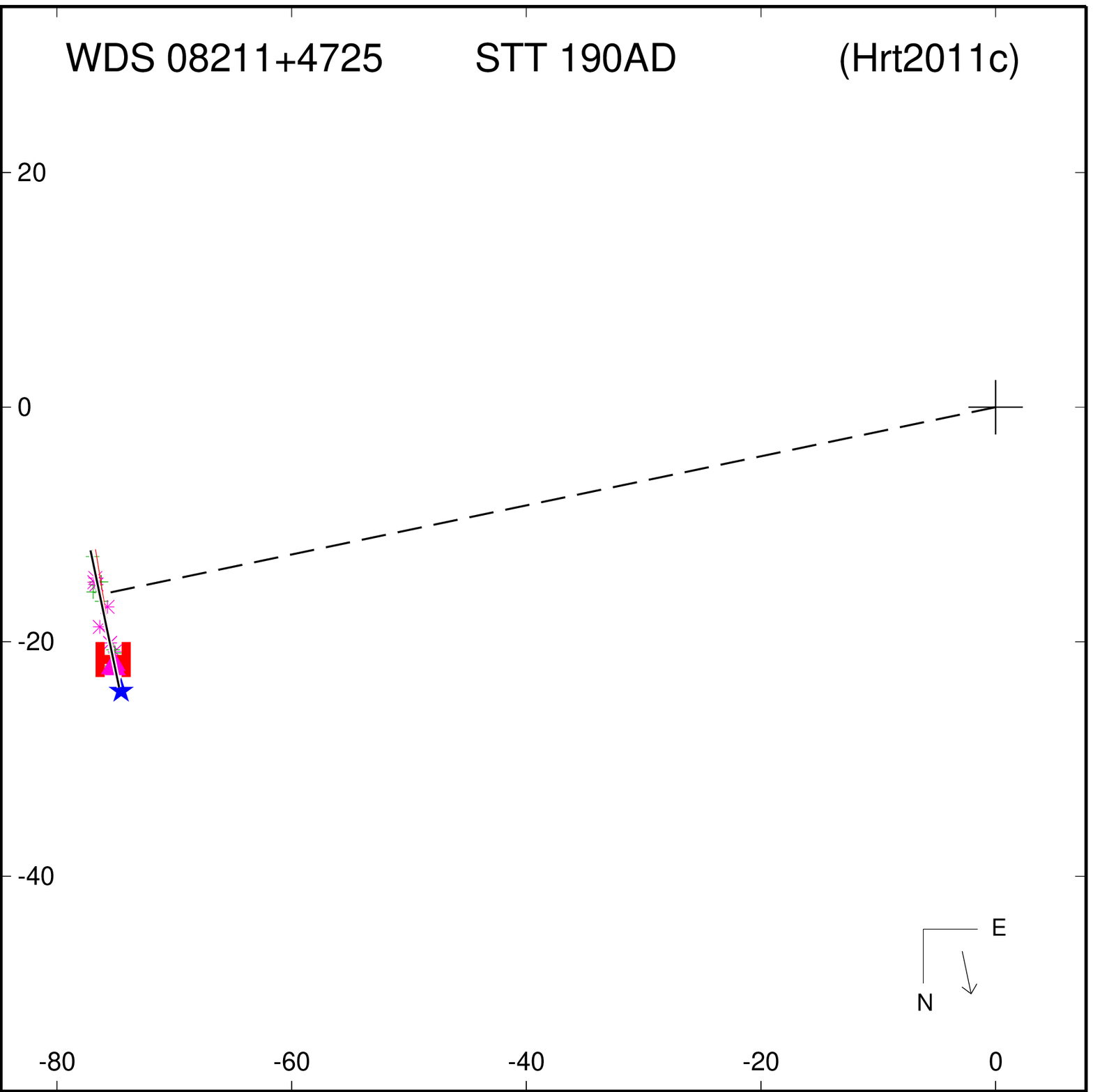}} 
\vskip 0.05in
{\epsfxsize 2.8in \epsffile{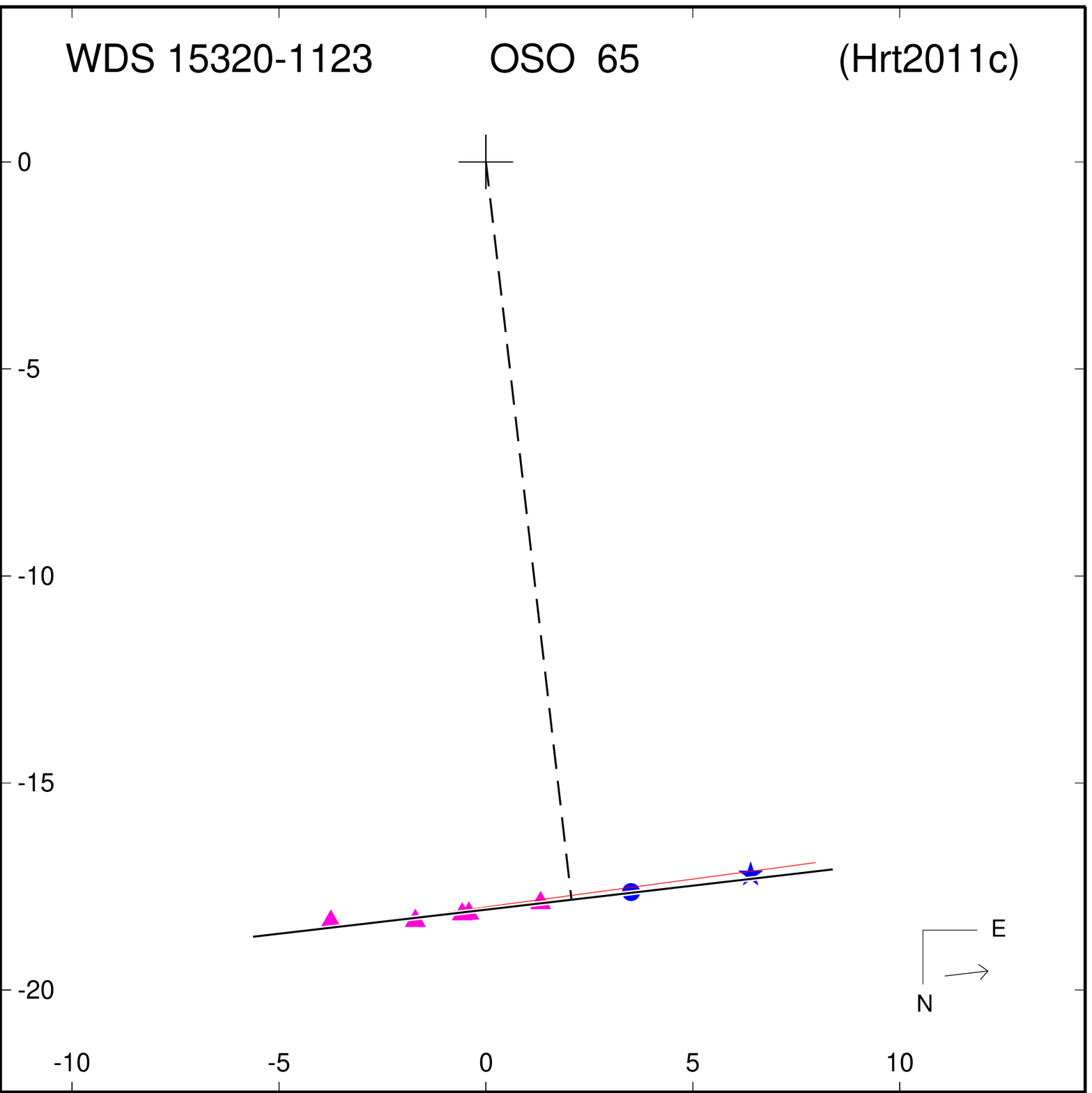} \epsfxsize 2.8in \epsffile{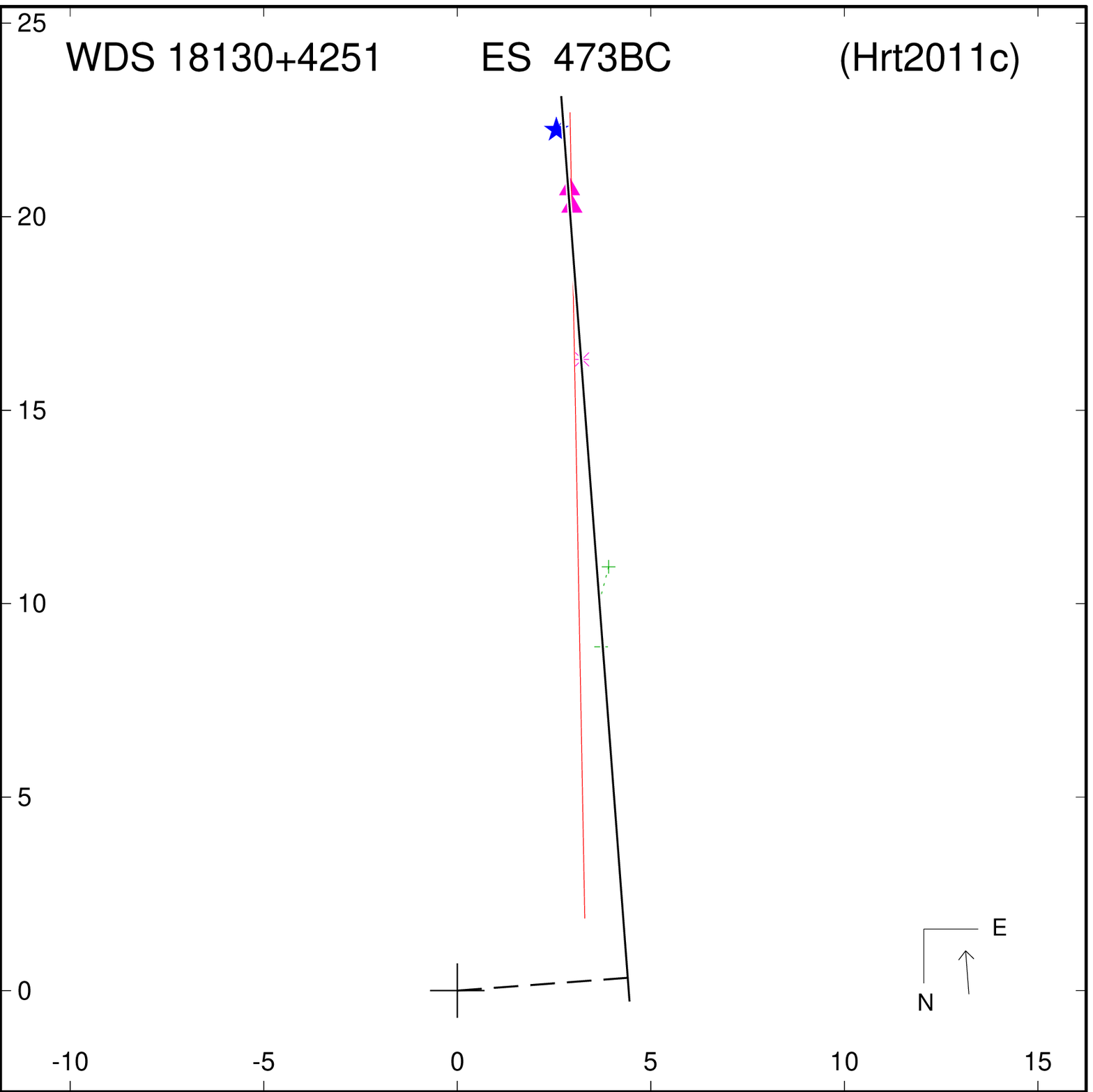}} 
\end{center}
\vskip -0.3in
\caption{\small New linear fits for the systems listed in Table 6 
and all data in the WDS database and Table 4. Symbols are the same 
as Figure 1. ``$O-C$" lines connect each measure to its predicted 
position along the linear solution (shown as a thick solid line). An
arrow in the lower right corner of each figure indicates the 
direction of motion. The scale, in arcseconds, is indicated on the 
left and bottom of each plot. When determined, cataloged proper motion
differences between these components is plotted as a red line.}
\end{figure}

\vskip 0.1in

Table 7 gives ephemerides for each orbit or linear solution over the
years 2018 through 2026, in two-year increments. Columns (1) and (2) 
are the same identifiers as in the previous tables, while columns 
(3+4), (5+6), ... (11+12) give predicted values of $\theta$ and 
$\rho$, respectively, for the years 2018.0,  2020.0, etc., through 
2026.0. 

Notes to individual systems follow:

{\bf 09210$+$3811 = STF1338AB} : Also known as HD 80441. Based on 
the period and semi-major axis of Table 5 and the parallax 
(23.44$\pm$1.08 mas; van Leeuwen 2007) the mass sum of this system 
is 1.66$\pm$0.33 \msun. This is lower than expected for a pair of F3
dwarfs. Using the more recent parallax from Gaia's DR2 (14.90$\pm$0.59
 mas; Gaia Collaboration et al.\ 2016, 2018) an even lower solution of
is 1.18$\pm$0.26 \msun is determined. If the spectral classification 
is approximately correct, an orbital solution of the same period with 
a semi-major axis about $\frac{1}{3}$ larger would produce an expected 
mass sum. While it has been 188 years since the first resolution 
(Struve 1837), only continued observation, over a long timebase, can 
make the orbital solution more definitive. The wider C component is 
optical.

{\bf 11035$+$5432 = A\phn\phn1590} : Also known as HD 95690. Based 
on the period and semi-major axis of Table 5 and the parallax 
(23.06$\pm$1.48 mas; van Leeuwen 2007) the mass sum of this system 
is 1.73$\pm$0.70 \msun. Using the more recent parallax from Gaia's 
DR2 (23.49$\pm$0.05 mas; Gaia Collaboration et al.\ 2016, 2018) a 
more precise result of 1.64$\pm$0.36 \msun is determined. Both seems 
reasonable for a K2V and its companion. 

\subsection{Double Stars Not Found}

Table 8 presents two systems which were observed but not detected. 
Possible reasons for nondetection include orbital or differential 
proper motion making the binary too close or too wide to resolve at 
the epoch of observation, a larger than expected $\Delta$m, 
incorrect pointing of the telescope, and misprints and/or errors in 
the original reporting paper. It is hoped that reporting these will 
encourage other double star astronomers to either provide 
corrections to the USNO observations or to verify the lack of 
detection.

\acknowledgements

This research has also made use of the SIMBAD database, operated at 
CDS, Strasbourg, France, NASA's Astrophysics Data System and made 
use of data from the European Space Agency (ESA) mission {\it Gaia} 
({\tt https://www.cosmos.esa.int/gaia}), processed by the {\it Gaia}
Data Processing and Analysis Consortium (DPAC, 
{\tt https://www.cosmos.esa.int/web/gaia/dpac/consortium}). Funding 
for the DPAC has been provided by national institutions, in 
particular the institutions participating in the {\it Gaia} 
Multilateral Agreement. 

The continued instrument maintenance by the USNO instrument shop, 
Gary Wieder, Chris Kilian and Phillip Eakens, makes the operation of 
a telescope of this vintage a true delight.


\begin{deluxetable}{lcl@{~}r@{~}lccc}
\tablenum{1}
\tablewidth{0pt}
\tablecaption{New Pairs}
\tablehead{
\colhead{Coordinates} & 
\colhead{WDS} &
\multicolumn{3}{c}{Discoverer} &
\multicolumn{2}{c}{Magnitudes} &
\colhead{Note}  \nl
\colhead{$\alpha,\delta$ (2000)} &
\colhead{Designation} &
\multicolumn{3}{c}{Designation} & 
\colhead{Primary} &
\colhead{Secondary} & 
\colhead{~}
}
\startdata
05\,22\,45.12 $+$53\,32\,09.6 & 05228$+$5332 & HJ  & 2258 & AC    &  9.59    &  14.0\phn    & 1   \nl
\enddata
\tablenotetext{~}{1 : AB is the historically known pair. This is a closer pair.}
\end{deluxetable}


\begin{deluxetable}{ll@{~}r@{~}lccclc}
\tablenum{2}
\tablewidth{0pt}
\tablecaption{Data Mining Measures for New Pairs}
\tablehead{
\colhead{WDS Desig.} & 
\multicolumn{3}{c}{Discovery} & 
\colhead{JY} &
\colhead{$\theta$} & 
\colhead{$\rho$} &
\colhead{Source} & 
\colhead{Reference} \nl
\colhead{$\alpha,\delta$ (2000)} &
\multicolumn{3}{c}{Designation} & 
\colhead{~} & 
\colhead{($^{\circ}$)} & 
\colhead{($''$)} &
\colhead{~} &
\colhead{or Note} \nl
}
\startdata
05228$+$5332 & HJ  & 2258 & AC & 1998.94\phn\phn    &     56.0 & \phn8.03\phn & 2MASS & 1                      \nl
             &     &      &    & 2003.053\phn       &     56.1 & \phn8.310    & UCAC4 & Zacharias et al.\ 2013 \nl
\enddata
\tablenotetext{~}{1 : Cutri et al.\ (2003), All-sky Release. See {\tt Vizier On-line Data Catalog: II/246}.}
\end{deluxetable}


\scriptsize
\begin{deluxetable}{ll@{~}r@{~}lrrrrrrr}
\label{tab:linear}
\tablenum{6}
\tablewidth{0pt}
\tablecaption{New Linear Elements}
\tablehead{
\colhead{WDS Desig.} & 
\multicolumn{3}{c}{Discoverer} &
\colhead{x$_0$} & 
\colhead{a$_{\rm x}$} & 
\colhead{y$_0$} &
\colhead{a$_{\rm y}$} &
\colhead{T$_0$} &
\colhead{$\rho_0$} & 
\colhead{$\theta_0$} \nl
\colhead{$\alpha,\delta$ (2000)} &
\multicolumn{3}{c}{Designation} & 
\colhead{($''$)} & 
\colhead{($''$/yr)} & 
\colhead{($''$)} &
\colhead{($''$/yr)} &
\colhead{(yr)} &
\colhead{($''$)} & 
\colhead{($\deg$)} \nl
\colhead{(1)} & 
\multicolumn{3}{c}{(2)} & 
\colhead{(3)} & 
\colhead{(4)} & 
\colhead{(5)} &
\colhead{(6)} &
\colhead{(7)} &
\colhead{(8)} & 
\colhead{(9)} \nl
}
\startdata
03313$-$0542 & HLD    &   65 &      & $-$1.387066 &    0.018936 &  $-$3.307647 & $-$0.007941 & 1799.833 &  3.587 &  337.25 \nl
05313$-$1834 & B      & 2588 &      & $-$4.232138 & $-$0.078973 &     3.165629 & $-$0.105579 & 1876.193 &  5.285 &  233.20 \nl
07377$-$1330 & J      & 2840 &      & $-$5.459991 & $-$0.043010 &  $-$4.515467 &    0.052006 & 1951.682 &  7.085 &  309.59 \nl
08211$+$4725 & STT    &  190 & AD   &$-$76.353394 &    0.015329 & $-$15.976615 & $-$0.073260 & 1914.322 & 78.007 &  281.82 \nl
15320$-$1123 & OSO    &   65 &      &    2.068836 &    0.385139 & $-$17.819626 &    0.044714 & 2006.111 & 17.939 &    6.62 \nl
18130$+$4251 & ES     &  473 & BC   &    4.403793 & $-$0.009231 &     0.332003 &    01224365 & 1836.467 &  4.416 &   94.31 \nl
\enddata
\end{deluxetable}


\scriptsize
\begin{deluxetable}{ll@{~}r@{~}cccclccc}
\tablenum{8}
\tablewidth{0pt}
\tablecaption{Double Stars Not Found}
\tablehead{
\colhead{WDS Desig.} & 
\multicolumn{3}{c}{Discoverer} &
\multicolumn{4}{c}{Most Recent Published Observation} &
\multicolumn{2}{c}{Published Magnitudes} &
\colhead{Notes} \nl
\colhead{$\alpha,\delta$ (2000)} &
\multicolumn{3}{c}{Designation} & 
\colhead{Date} &
\colhead{$\theta$} &
\colhead{$\rho$} &
\colhead{Reference} &
\colhead{Primary} &
\colhead{Secondary} &
\colhead{~} \nl
\colhead{~} &
\multicolumn{2}{c}{~} & 
\colhead{~} &
\colhead{(${\circ}$)} & 
\colhead{($''$)} &
\colhead{~} &
\colhead{~} &
\colhead{~} &
\colhead{~}
}
\startdata
06216$-$2002 & HDO &   82 & AB & 1869 &    180 &    12.0 & Winlock \& Pickering 1882  & \phn9.\phn & 11.\phn & 1 \nl
14362$-$0636 & BRT &  455 &    & 1911 & \phn87 & \phn4.0 & Barton 1931                &    10.9    & 11.0    & 2 \nl
\enddata
\tablenotetext{~}{1: Triple system. While AC has six measure, AB has never been confirmed. The direction
of the component is only approximately given as $``$south". Maybe the proper motion of B moved it out of the 
field of view of the A component when AC was confirmed in 1981 (Tobal 2005).}
\tablenotetext{~}{2: Remeasured more accurately in the AC reanalysis (Urban et al.\ 1998), but this may be nothing 
more than a more precise measure of a plate flaw.}
\end{deluxetable}

\end{document}